\pdfoutput=1

\documentclass[11pt]{article}

\usepackage[]{acl}

\usepackage{times}
\usepackage{latexsym}
\usepackage{cite}
\usepackage{natbib}

\usepackage[T1]{fontenc}

\usepackage{microtype}

\usepackage{amsmath}
\usepackage{amsfonts}
\usepackage{graphicx}  
\usepackage{xspace}
\usepackage{eufrak}
\usepackage{bm}
\usepackage{multirow}
\usepackage{booktabs}
\usepackage{subfig}
\usepackage[utf8]{inputenc}

\usepackage{cleveref}
\usepackage{enumitem}
\usepackage{adjustbox}

\usepackage{xargs}
\usepackage{etoolbox}
\usepackage[flushleft]{threeparttable}

\usepackage[colorinlistoftodos,prependcaption,textsize=tiny]{todonotes}
\usepackage[group-separator={,}, group-minimum-digits=4]{siunitx}
\usepackage{comment}

\setlist{nosep}

\crefformat{section}{\S~#2#1#3} 
\crefformat{subsection}{\S~#2#1#3}
\crefformat{subsubsection}{\S~#2#1#3}

\newcommand{\safedefine}[2]{
  \ifdef{#1}{\renewcommand{#1}{#2}}{\newcommand{#1}{#2}}
}
\makeatletter
\newcommand\footnoteref[1]{\protected@xdef\@thefnmark{\ref{#1}}\@footnotemark}
\makeatother

\newcommand{\ours}{\textsc{TrENC}\xspace}
\newcommand{\dataset}{\textsc{WebPT}\xspace}
\newcommand*{\rom}[1]{\uppercase\expandafter{\romannumeral #1}}

\newcommandx{\chao}[2][1=]{\todo[linecolor=red,backgroundcolor=red!25,bordercolor=red,#1]{#2}}
\newcommandx{\yli}[2][1=]{\todo[linecolor=blue,backgroundcolor=blue!25,bordercolor=blue,#1]{#2}}
\newcommandx{\prashant}[2][1=]{\todo[linecolor=green,backgroundcolor=green!25,bordercolor=green,#1]{#2}}
\newcommandx{\colin}[2][1=]{\todo[linecolor=purple,backgroundcolor=purple!25,bordercolor=red,#1]{#2}}
\newcommandx{\improvement}[2][1=]{\todo[linecolor=yellow,backgroundcolor=yellow!25,bordercolor=yellow,#1]{#2}}
\newcommandx{\thiswillnotshow}[2][1=]{\todo[disable,#1]{#2}}

\newcommand{\ie}{\emph{i.e.}\xspace} 
\newcommand{\Eg}{\emph{E.g.}\xspace} 
\newcommand{\wrt}{\emph{w.r.t.}\xspace}

\safedefine{\iff}{\emph{i.f.f.}\xspace}
\safedefine{\iid}{\emph{i.i.d.}\xspace}
\newcommand{\fone}{F\textsubscript{1}\xspace}
\newcommand{\vs}{\emph{vs.}\xspace}
\newcommand{\st}{\emph{s.t.}\xspace}

\newcommand{\dn}{$\downarrow$}

\newcommand{\ba}{\bm{a}}

\newcommand{\bc}{\bm{c}}
\newcommand{\bb}{\bm{b}}

\newcommand{\W}{\bm{W}}

\newcommand{\x}{\bm{x}}
\newcommand{\bi}{\bm{i}}

\newcommand{\s}{\bm{s}}
\newcommand{\g}{\bm{g}}
\newcommand{\bt}{\bm{t}}

\newcommand{\e}{\bm{e}}
\newcommand{\w}{\bm{w}}

\newcommand{\T}{\mathcal{T}}
\newcommand{\calS}{\mathcal{S}}
\newcommand{\calO}{\mathcal{O}}

\newcommand{\bH}{\bm{H}}

\newcommand{\R}{\mathbb{R}}

\newcommand{\N}{\mathcal{N}}
\newcommand{\M}{\bm{M}}

\newcommand{\dm}{{d_{\rm model}}}
\newcommand{\V}{\mathcal{V}}
\newcommand{\E}{\mathcal{E}}

\newcommand{\softmax}{\text{SoftMax}}

\newcommand{\tr}{\mathsf{T}}

\title{Extracting Shopping Interest-Related Product Types from the Web}

\author{
  Yinghao Li$^{1}$\Thanks{\noindent All work performed while interning at Amazon.} , Colin Lockard$^2$, Prashant Shiralkar$^2$, Chao Zhang$^1$\\
  $^1$Georgia Institute of Technology, Atlanta, USA\quad $^2$Amazon, Seattle, USA\\
  $^1$\texttt{$\{$yinghaoli,chaozhang$\}$@gatech.edu}\\
  $^2$\texttt{$\{$clockard,shiralp$\}$@amazon.com}
}

\begin{document}
\maketitle

\begin{abstract}
  Recommending a diversity of product types (PTs) is important for a good shopping experience when customers are looking for products around their high-level shopping interests (SIs) such as hiking.
  However, the SI-PT connection is typically absent in e-commerce product catalogs and expensive to construct manually due to the volume of potential SIs, which prevents us from establishing a recommender with easily accessible knowledge systems.
  To establish such connections, we propose to extract PTs from the Web pages containing hand-crafted PT recommendations for SIs.
  The extraction task is formulated as binary HTML node classification given the general observation that an HTML node in our target Web pages can present one and only one PT phrase.
  Accordingly, we introduce \ours, which stands for \textbf{Tr}ee-\textbf{Tr}ansformer \textbf{Enc}oders for \textbf{N}ode \textbf{C}lassification.
  It improves the inter-node dependency modeling with modified attention mechanisms that preserve the long-term sibling and ancestor-descendant relations.
  \ours also injects SI into node features for better semantic representation.
  Trained on pages regarding limited SIs, \ours is ready to be applied to other unobserved interests.
  Experiments on our manually constructed dataset, \dataset, show that \ours outperforms the best baseline model by \num{2.37} \fone points in the zero-shot setup.
  The performance indicates the feasibility of constructing SI-PT relations and using them to power downstream applications such as search and recommendation.
\end{abstract}

\section{Introduction}
\label{sec:intro}

\begin{figure}[!t]
    \centerline{\includegraphics[width = 0.46\textwidth]{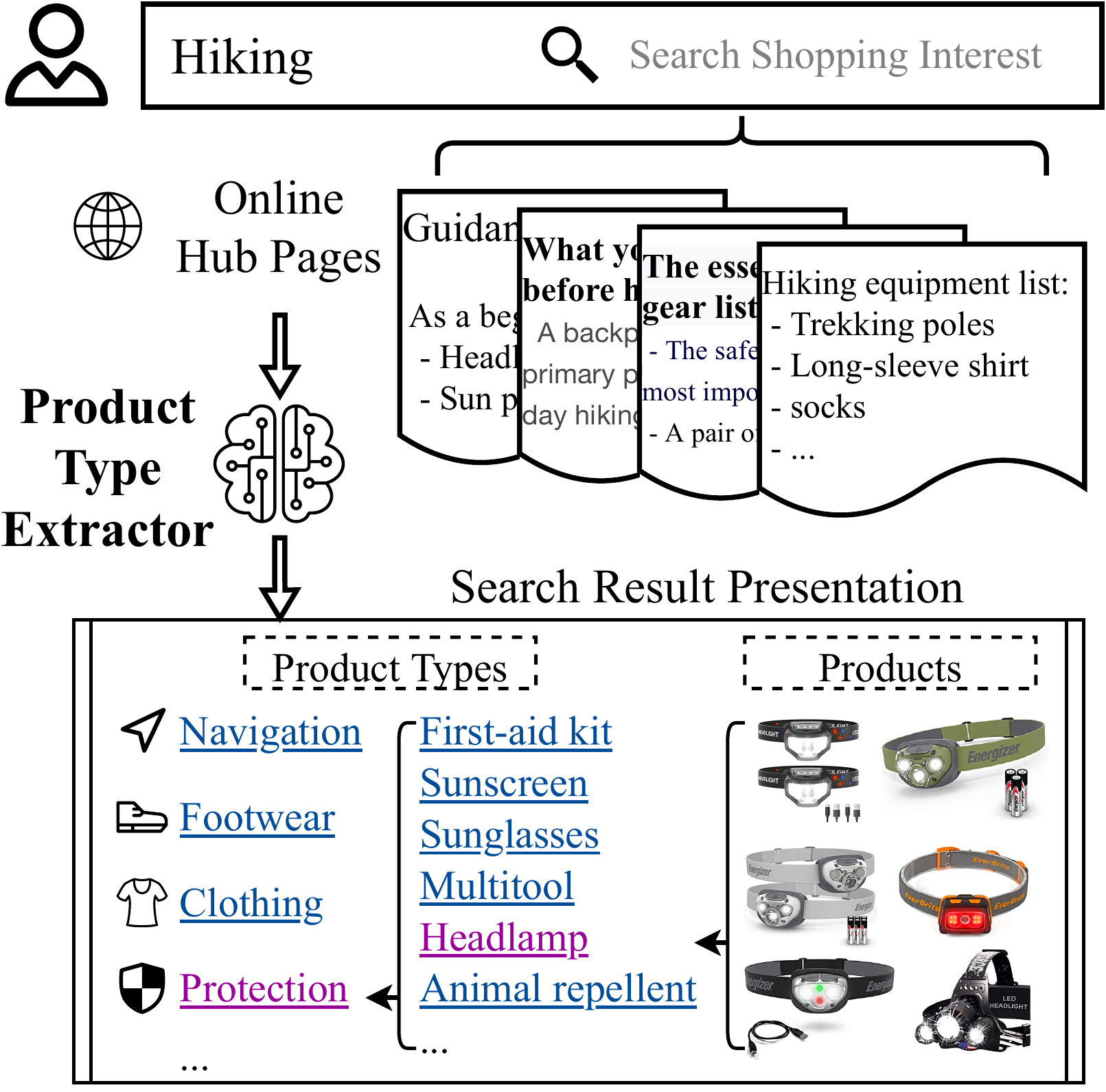}}
    \caption{
        Example of the system and results to deliver as a response to searching for the SI ``hiking''.
    }
    \label{fig:1-motivation}
\end{figure}

Customers of e-commerce websites fall in various stages of the purchase funnel\footnote{\href{https://en.wikipedia.org/wiki/Purchase\_funnel}{https://en.wikipedia.org/wiki/Purchase\_funnel}} in their journey to purchase specific products.
While lower-funnel customers target specific products or product categories, a customer in the middle to upper funnel only has vague shopping interests (SIs) and requires additional guidance to determine the right products to purchase.
Existing e-commerce websites are limited today in their ability to assist them in this kind of interest-oriented shopping.
For example, a customer searching for \emph{COVID-19 crisis} gets top results showing product types (PTs) such as books and test kits, while missing other essential categories such as the face mask, thermometer, or medicine.
Moreover, the search result is a random assortment of products, without a clear organization that helps upper-funnel customers discover products within relevant categories.

The main problem is the concept of ``shopping interest'' is generally absent in e-commerce catalogs, which makes it difficult to directly establish the SI-PT connections and give corresponding recommendations.
To circumvent such system limitations, customers today are accustomed to researching their products on hand-curated ``hub Web pages''\footnote{\Eg, \href{https://hikinginthesmokys.com/hiking-checklist/}{hikinginthesmokys.com/hiking-checklist} for hiking.}, each related to an SI and presenting PT suggestions as organized lists, before returning to e-commerce websites.
This stretches the total time spent on a purchase.
We aim to find SI-related PTs directly on the e-commerce website, reducing customer effort for all their interest-oriented needs.
Figure~\ref{fig:1-motivation} shows the desired search experience.

The first step to this end is collecting hub pages, which is realized by querying Google Search with automatically selected prompts (\cref{appsec:dataset.construction.details}).
The rest of the paper focuses on PT extraction from the HTML pages, which presents several challenges.
First, hub websites are heterogeneous in their format and terminology, with PTs often interspersed among long descriptive paragraphs, making it challenging for any solution designed for one or a few websites to work well for others.
Second, our page collection approach assumes that all PTs presented on a page are related to the same SI, which may not hold true in practice, requiring us to filter out irrelevant PTs.
Finally, our goal to find PTs for a wide range of SIs motivates us to consider a zero-shot learning setup \citep{Xian.2019.Zero.Shot.Learning} \wrt SIs, to generalize to interests not seen during training.

Representing an HTML document by a Document Object Model (DOM) tree whose nodes are HTML tags with text sequences, we formulate PT extraction as a node classification task that entails checking whether its text sequence represents a PT phrase.
It is based on the empirical discovery that in our collected hub pages, a PT phrase generally occupies a single DOM node within a coherent group of enumerated HTML elements such as section titles or bullet points, where knowing one PT phrase suggests the potential presence of other PT phrases in the neighboring elements (Figure~\ref{subfig:dom.tree}).
Node classification emphasizes learning inter-node structural dependencies rather than intra-node token interactions, which results in better generalization to a wide variety of HTML structures.

Due to the absence of a dedicated DOM tree encoding method, we propose \ours (\textbf{Tr}ee-\textbf{Tr}ansformer \textbf{Enc}oders for \textbf{N}ode \textbf{C}lassification) to fill in the blanks.
Adapted from the Transformer \citep{Vaswani.2017.Transformer}, \ours incorporates ancestor-descendant and sibling node relations using modified self-attention mechanisms and positional embeddings that are suited to the unique DOM node arrangement of the rendered hub pages.
The ancestor-descendant relation provides relative structural information between nodes in the DOM node hierarchy, whereas the sibling relation tracks the semantical connection among sibling nodes.
The modified attention mechanisms reconstruct the tree architecture from the linearized input nodes and facilitate long-term dependency modeling.
To capture the relevance between an SI and a node, we leverage a gating network to dynamically integrate SI semantics with a node's textual semantics, which generalizes \ours to unseen SIs.

Evaluated on our dataset \dataset with \num{453} Web pages covering \num{95} interests, \ours achieves $2.37$ absolute \fone performance gain over the strongest baseline method.
Our contributions include
\begin{itemize}[leftmargin=*]
    \item a novel and practical research topic of product type extraction from the Web pages associated with a given shopping interest;
    \item \ours, a Transformer encoder-based model with structural attention mechanisms for recovering the DOM tree architecture from the node sequence to promote classification;
    \item a dataset \dataset, and comprehensive evaluations of graph encoding techniques to verify the effectiveness of our model design.
\end{itemize}
The dataset is made publicly accessible at \href{https://github.com/Yinghao-Li/WebIE}{https://github.com/Yinghao-Li/WebIE} to promote future research.

\section{Related Works}
\label{sec:related.works}

\paragraph{Web Information Extraction}

Information extraction from the semi-structured Web data is a long-studied topic \citep{Chang.2006.Survey.WebIE, Banko.2007.OpenIE, Sleiman.2013.Survey.Regioin.Extractor}.
The works most relevant to ours are those on product attribute extraction \citep{Zheng.2018.OpenTag, xu.2019.scaling, lockard.2020.zeroshotceres, Zhou.2021.Simplified.DOM, Wang.2022.WebFormer, Deng.2022.DOM-LM}.
For example, \citet{Zheng.2018.OpenTag} train a BiLSTM-CRF network \citep{Huang.2015.BiLSTMCRF} for each attribute to locate its corresponding values on text sequences.
\citet{xu.2019.scaling} scale it up by injecting the attribute name into the network as an attention objective.
\citet{Wang.2022.WebFormer} encode the DOM tree with graph attention network \citep{vel.2018.GAT} to incorporate the dependencies between nodes.
However, attribute extraction is different from our PT extraction task at two major points.
First, attributes are typically extracted from product detail pages, each of which mentions multiple attributes; and the attribute name-value pairs cluster around titles, bullet points and descriptions.
In contrast, a hub page generally focuses on a single SI, with PTs scattered throughout the page.
Unlike attribute extraction approaches that limit the searching scope to certain regions, the characteristics of hub pages require us to consider a page holistically instead of a small part.
Second, attribute extraction is performed as token-level entity recognition in previous works, while PT extraction requires a node-level classification, which prevents approaches for the former from being directly applied to the latter.
To our best knowledge, no applicable DOM node classification or similar dataset exists in openly available benchmarks such as OGB \citep{Hu.2020.OGB}.

\paragraph{Graph Transformers}


Recently, graph neural networks (GNNs) such as the graph convolutional network (GCN, \citealp{Kipf.2017.GCN}) and graph attention network (GAT, \citealp{vel.2018.GAT, Brody.2022.GATv2}) have dominated the graph encoding research.
But some works try to model graphs using Transformers \citep{Vijay.2020.a.generalization, Maziarka.2020.Molecule.Attention.Transformer, Ying.2021.Graphormer, Park.2022.Deformable.Graph.Transformer, wu.2022.nodeformer}, to which our work is more related.
For example, \citet{Maziarka.2020.Molecule.Attention.Transformer} add inter-atomic distances into the self-attention heads to parameterize the molecular graph structure.
Also targeting molecules, Graphormer \citep{Ying.2021.Graphormer} takes a step further and introduces centrality encoding, edge encoding and spacial encoding to evaluate the atom importance and capture the edge and graph structure.
\citet{Park.2022.Deformable.Graph.Transformer} and \citet{wu.2022.nodeformer} extend Transformers to knowledge graphs with partial message-passing tricks.
Although applicable, the hierarchical and acyclic nature of DOM trees is different from the graphs for which the approaches were designed.
Directly applying them to DOM trees leads to sub-optimal performance, as shown in \cref{sec:experiment}.

\section{Problem Setup}
\label{sec:problem.def}
We possess the DOM tree of a Web page associated with a given shopping interest $C$.
The DOM tree can be represented by a set of nodes $\V = \{V_1, V_2, \dots, V_{|\V|}\}$ as well as a set of edges $\E = \{E_1, E_2, \dots, E_{|\E|}\}$ that connect the parent and children nodes.
$|\V|$ and $|\E|$ are the sizes of node and edge sets respectively.
We aim to design a binary node classifier $f: \V \cup \E \cup \{C\} \mapsto \{0,1\}^{|\V|}$ to judge whether the text sequence in each node is a phrase representing a product type.
The nodes with positive labels are referred to as ``PT nodes'' and the labels are denoted by $y_m=1, m\in 1:|\V|$.
We focus our discussion on \emph{one} DOM tree and use $m\in 1:|\V|$ as its node index.

\begin{figure}[!t]
    \centerline{\includegraphics[width = 0.48\textwidth]{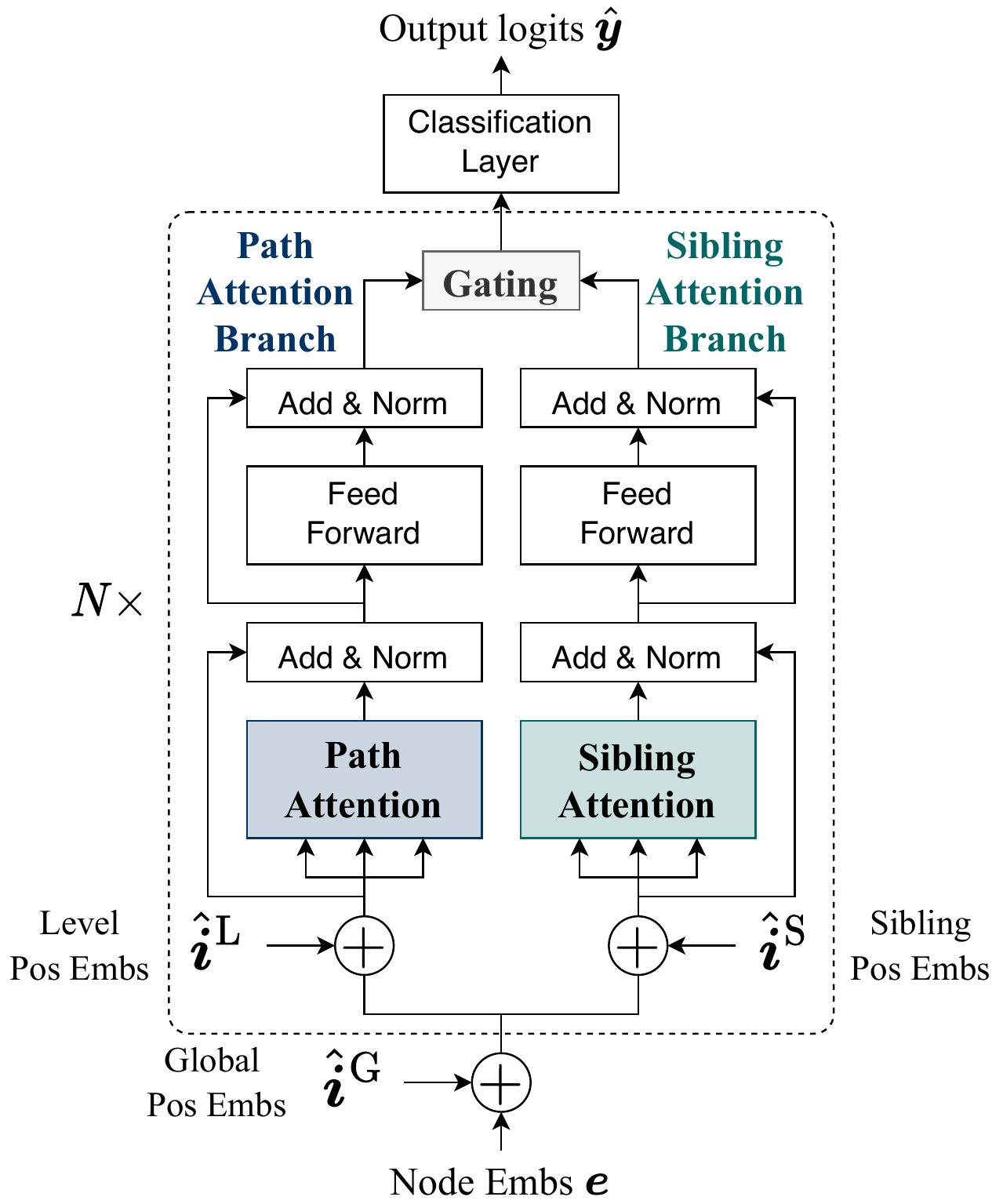}}
    \caption{
        Model structure.
        It adopts path and sibling attention mechanisms to model the ancestor-descendant and sibling relationships among DOM nodes.
    }
    \label{fig:s4-model-architecture}
\end{figure}

\begin{figure*}[tbp]
  \centering{
    \subfloat[Example of a simple DOM tree and corresponding positional indices]{
      \label{subfig:dom.tree}
      \includegraphics[height=1.9 in]{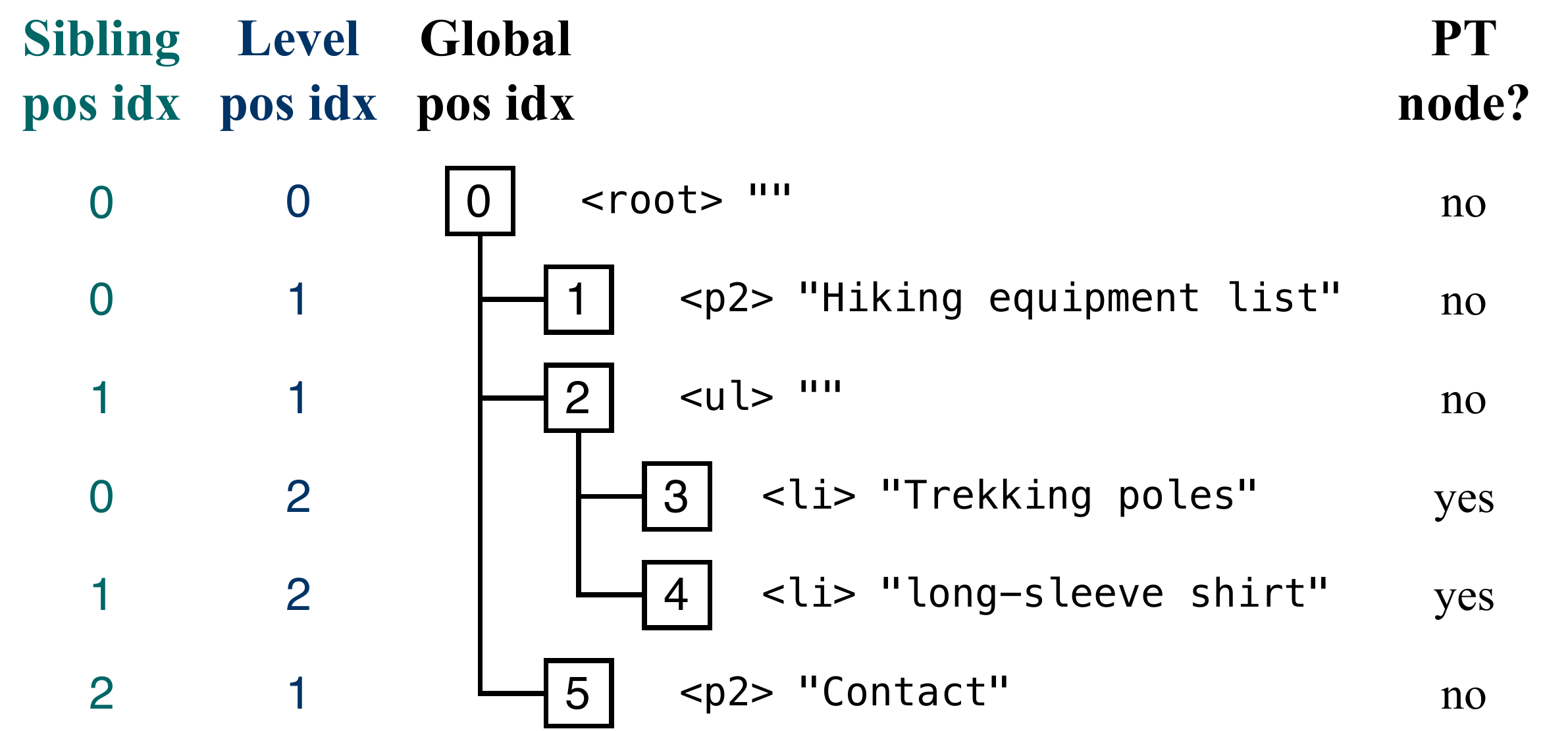}
    }
    \subfloat[Node sets and attention masks]{
      \label{subfig:attn.masks}
      \includegraphics[height=1.9in]{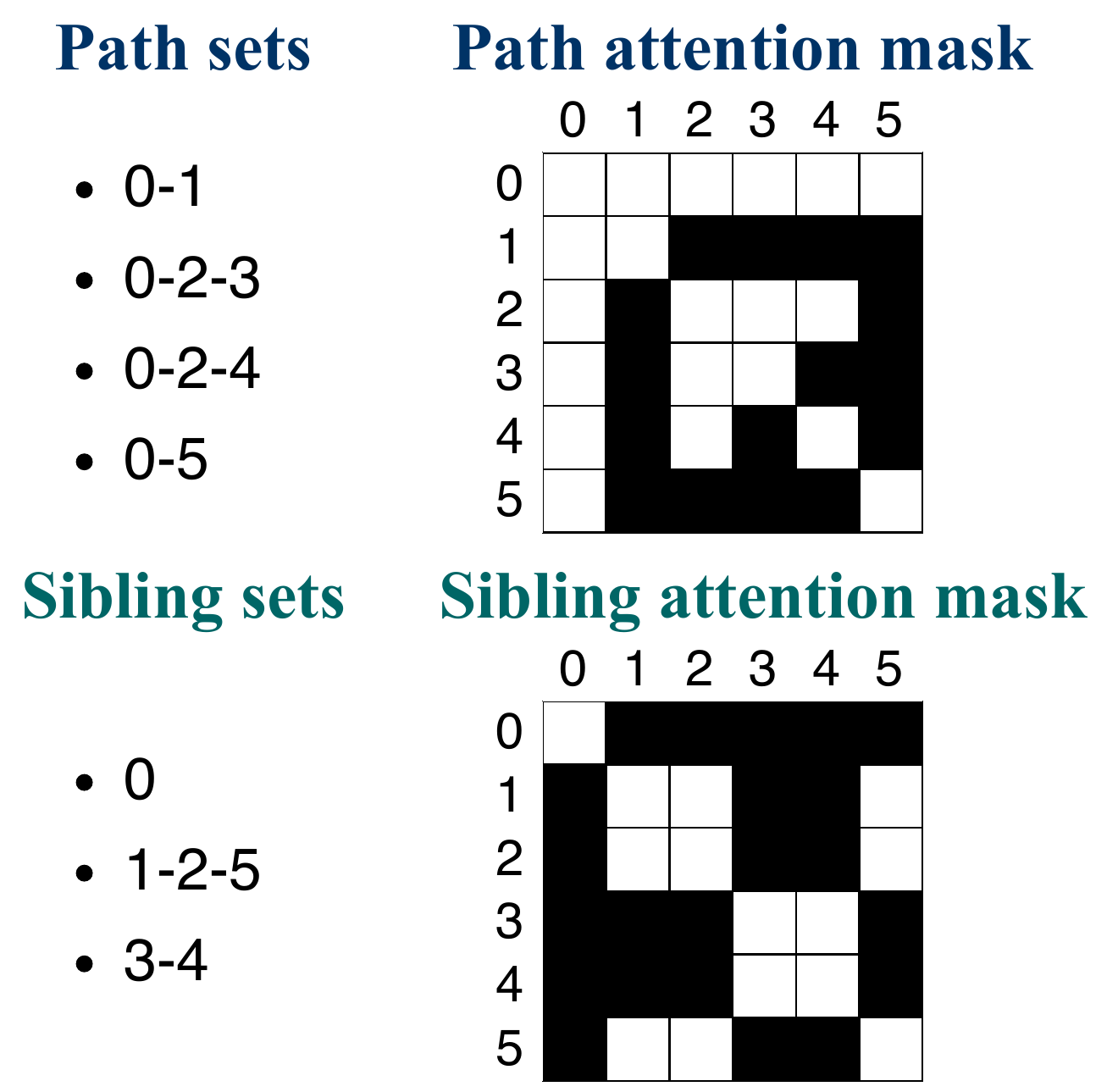}
    }
  }
  \caption{
    An example of the model inputs, including the DOM tree components, positional indices and attention masks.
    The path sets and sibling sets in \protect\subref{subfig:attn.masks} are defined by the global positional indices in \protect\subref{subfig:dom.tree}.
    In the attention masks, white elements have values $0$ and the black ones are $-\infty$.
  }
  \label{fig:s3-inputs}
\end{figure*}

\section{Method}
\label{sec:method}

%
We propose \ours to model the DOM tree of hub Web pages for PT extraction.
Figure~\ref{fig:s4-model-architecture} shows the model architecture.
We treat the problem as a DOM node classification task that entails detecting whether its textual sequence defines a PT phrase.
We first create a node representation that integrates three basic signals of a node that may be indicative of a PT (\cref{subsec:input.features}).
We then adapt the Transformer architecture by adding two attention mechanisms, namely path attention and sibling attention, that allow capturing of inter-node dependencies presented by their HTML structure (\cref{subsubsec:structural.attentions}).
We also include three kinds of positional encodings that assist the attention layers with the node's unique positional information within the DOM tree (\cref{subsubsec:positional.encodings}).
Finally, we integrate the outputs from the path and sibling attention layers, which are used in a classification layer to predict node labels (\cref{subsubsec:treeformer.layers}).
The implementation details are in \cref{appsubsec:imp.details.treeformer}.



\subsection{Node Features}
\label{subsec:input.features}

Besides the SI $C$ associated with the tree, we consider two features for each node $V_m$:
1) its HTML tag $t_m \in \T$ where $\T$ is a finite tag set;
and 2) the text sequence $\calS_m=\{w_{m, 1}, w_{m, 2}, \dots, w_{m, |\calS_m|}\}$, where $|\calS_m|$ is the length and $w$ is the token.

\paragraph{HTML Tag}

HTML tags are a finite vocabulary of keywords that define how browsers display their content.
Specifically, they convey the semantic nature of their enclosed content.
For example, \texttt{<p>} denotes a paragraph, while \texttt{<ul>} represents a list.
Based on the observation that some tags tend to contain PT phrases more than others, we capture the tag information as a distinct structural feature and encode $t_m$ with a vector $\bt_m\in\R^{\dm}$ using an embedding layer.
Here, $\dm$ is the model dimensionality as in Transformers.

\paragraph{Text Sequence}
Text sequences convey the semantic character of an HTML document.
In addition to directly indicating a PT phrase, they can also serve as useful contextual information about the neighboring nodes' propensity to contain a PT phrase.
For example, a node stating ``Essentials for camping'' is a clear indicator that what follows is likely a set of camping-related PT nodes.

We leverage the power of pre-trained language models (PLMs) such as BERT \citep{Devlin.2019.BERT} to properly encode their semantics.
For a given sequence, BERT generates an embedding $\w_{m, i}\in\R^{d_{\rm BERT}}$ for each token $w_{m, i}, i\in 1:|\calS_m|$, besides two special tokens $w_{m, 0}$ and $w_{m, |\calS_m|+1}$ representing the start and end of the sequence.
We derive the sequence embedding $\s_m \in \R^\dm$ by taking an average of all the token embeddings and passing it through a feed-forward network (FFN) layer:
\begin{equation}
    \label{eq:text.emb}
    \s_m = \W^{\rm seq}({\rm GELU}(\frac{1}{|\calS_m|+2}\sum_{i=0}^{|\calS_m|+1}\w_{m, i})),
\end{equation}
where $\W^{\rm seq}\in\R^{\dm \times d_{\rm BERT}}$ are parameters.

\paragraph{Shopping Interest}

Although we assume that a DOM tree is associated with only one SI $C$, in rare cases this assumption does not hold.
We are thereby motivated to capture the relevance between a node and the interest.
Accordingly, we incorporate $C$ with an embedding vector $\bc\in\R^\dm$ in a similar manner as that for node text sequence \eqref{eq:text.emb}, and let the model learn the relevance between $C$ and related PTs to rule out any false positive cases.

\paragraph{Feature Integration}

We integrate node features into the node embedding $\e_m\in\R^\dm$ in two steps to honor the distinctiveness between the structural feature $\bt_m$ and semantic features $\s_m$ and $\bc$.

First, we merge the semantic features.
Since different nodes have differing levels of correlations with the interest, we use gating vectors \citep{Hochreiter.1997.LSTM} to automatically control how much interest embeddings $\bc$ should be integrated into the sequence embedding $\s_m$.
We calculate the weights $\g$ as:
\begin{equation}
    \label{eq:gating}
    \g(\x_1, \x_2) = \sigma(\W_1\x_1 + \W_2\x_2 + \bb),
\end{equation}
where $\x_1$ and $\x_2$ are feature vectors; $\W_1$ and $\W_2$ are trainable square matrices; $\bb$ is the bias, and $\sigma$ is the sigmoid function.
With \eqref{eq:gating}, the updated sequence embedding vector becomes
\begin{equation*}
    \s'_m = \g(\bc, \s_m)\odot\bc + \s_m,
\end{equation*}
where $\odot$ is the element-wise product.

Then, we integrate the semantic and structural embeddings using concatenation followed by an FFN layer to maintain the embedding dimensionality.
The integrated node embedding $\e_m$ is
\begin{equation*}
    \e_m = \W^{\rm emb}[{\s'_m}^\tr; \bt_m^\tr]^\tr,
\end{equation*}
where $[\cdot;\cdot]$ represents vector concatenation and $\W^{\rm emb}\in\R^{\dm \times 2\dm}$ is an FFN layer.



\subsection{\ours Architecture}
\label{subsec:model.architecture}

Compared with conventional GNNs that generally aggregate only $1$-hop neighboring messages in each layer, Transformers are better at tracking long-term dependencies.
However, applying the Transformer encoder to DOM trees as is can lead us astray because it is not designed to naturally accommodate the hierarchical structure of a tree.
To address this limitation, we adapt the Transformer architecture by adding \emph{structural attention} mechanisms with \emph{node positional encodings} to better encode unique information within the DOM trees with the existing abilities of the Transformer architecture.

\subsubsection{Structural Attentions}
\label{subsubsec:structural.attentions}


The DOM tree structure presents two kinds of relations that convey how nodes are related.
The ancestor-descendant relation, represented by the edges $\E$, conveys the granular nature of a node (high or low) within the DOM hierarchy.
The sibling relation between nodes conveys how they semantically represent a coherent group, as shown in Figure~\ref{subfig:dom.tree}.
We incorporate these relationships via structural attention mechanisms, namely path attention and sibling attention.
Correspondingly, we represent these two views of the DOM tree by two types of node sets: \textit{path node sets} and \textit{sibling node sets}.
A path set $\N^{\rm P} \subset \V$ is the ordered collection of all nodes in an HTML path, from the root node to an arbitrary node, as illustrated in Figure~\ref{subfig:attn.masks}.
A sibling set $\N^{\rm S} \subset \V$ consists of the immediate children of a non-leaf node.
Thereupon, we develop \emph{path} and \emph{sibling attention} mechanisms, as described below, to explore the potential of modeling tree structures with Transformers.

\paragraph{Path Attention}

The path attention mechanism captures the granularity of a node $V_m$ within the DOM tree, which carries useful information about the node's tendency to present a PT phrase.
It limits the attention target of a DOM node to its ancestors or descendants only, echoing the edges $\E$ that define the DOM tree structure.
Path node sets help define an attention mask toward this purpose by leaving out all ``off-path'' elements during the self-attention message passing operation.

Suppose the input is $\bH^{\rm P}\in\R^{|\V|\times\dm}$, in each attention head, the path attention scores $\ba^{\rm P}_m\in (0,1)^{1\times|\V|}$ of $V_m$ attending to all DOM nodes are
\begin{equation}
    \label{eq:path.attn.weight}
    \ba^{\rm P}_m = \softmax ( \frac{ \bH^{\rm P}_m\W^{\rm Q}(\bH^{\rm P}{\W^{\rm K}})^\tr}{\sqrt{d_k}} + \M_m^{\rm P} ).
\end{equation}
Here $\W \in\R^{\dm \times d_k}$ are the FFN layers that map the latent features to the reduced $d_k$-dimensional single-head attention space, as in \citep{Vaswani.2017.Transformer}.
$\M^{\rm P}\in\{0,-\infty\}^{|\V|\times |\V|}$ is the path attention mask as shown in Figure~\ref{subfig:attn.masks}.
$\forall u,v\in 1:|\V|$,
\begin{equation}
    \label{eq:path.attn.mask}
    M_{u,v}^{\rm P} =
    \begin{cases}
        0, & \exists \N^{\rm P}\ \st\ V_u \in \N^{\rm P}, V_v \in \N^{\rm P}; \\
        -\infty, & \text{otherwise}.
    \end{cases}
\end{equation}
$\ba^{\rm P}_m$ has non-zero values at positions corresponding to $V_m$'s ancestors or descendants.
The single-head attention output of $V_m$ becomes
\begin{equation}
  \label{eq:path.attn}
  {\rm Attn}_m^{\rm P} = \ba^{\rm P}_m \bH^{\rm P}\W^{\rm V}.
\end{equation}
The rest of the architecture such as the layer norm and the residual connection is the same as in the Transformer and thus is omitted.

\paragraph{Sibling Attention}

Although sibling relations are not described by the edges $\E$, encoding them can provide a useful contextual signal based on the observation that sibling PT phrases often form a group.
Accordingly, analogous to path attention, we develop sibling attention by imposing an attention mask $\M^{\rm S}$, which forces a node to focus only on its siblings via self-attention.
The sibling node set $\N^{\rm S}$ helps define the mask.
%
Its calculation is identical to \eqref{eq:path.attn.weight}--\eqref{eq:path.attn}, except that the variables are superscripted by sibling ``$\cdot^{\rm S}$'' instead of path ``$\cdot^{\rm P}$''.

\subsubsection{Node Positional Encodings}
\label{subsubsec:positional.encodings}

Different from graphs, a DOM tree is acyclic and heterogeneous; the order of nodes influences their relations and how the elements are rendered.
As Transformers do not encode such node order, positional embeddings are critical to capture such positioning. \citep{Yun.2020.Are.Transformers}.
We consider three types of absolute indices: \emph{global}, \emph{level} and \emph{sibling} positional indices, as shown in Figure~\ref{subfig:dom.tree}.
The global positional index $i^{\rm G}_m$ represents the position of each node in the tree in the depth-first order.
It helps \ours understand how the nodes are organized in the rendered HTML pages.
The level index $i^{\rm L}_m$ and sibling index $i^{\rm S}_m$ on the other hand are developed to assist the path and sibling attentions.
$i^{\rm L}_m$ describes the level or depth of a node, to help distinguish a parent from its children during the path attention,
while the $i^{\rm S}_m$ captures the relative order among siblings within the sibling attention.

We encode positional indices by first applying sinusoid functions \citep{Vaswani.2017.Transformer} to convert them to vectors $\bi^{\rm G}_m, \bi^{\rm L}_m, \bi^{\rm S}_m \in [0,1]^{\dm}$, followed by applying an affine transformation that maps each of them into distinct latent spaces:
\begin{equation*}
  \hat{\bi}^{\rm G}_m = \W^{\rm G}\bi^{\rm G}_m;  \quad \hat{\bi}^{\rm L}_m = \W^{\rm L}\bi^{\rm L}_m;  \quad \hat{\bi}^{\rm S}_m = \W^{\rm S}\bi^{\rm S}_m,
\end{equation*}
where $\W\in \R^{\dm \times \dm}$ are FFN parameters.

\subsubsection{\ours Layers}
\label{subsubsec:treeformer.layers}

In each layer, the path and sibling signals are modeled by two parallel branches, which are identical except for the positional embeddings and attention mechanisms (Figure~\ref{fig:s4-model-architecture}).
Denoting the input feature of layer $l$ by $\bH^{(l)}\in\R^{|\V| \times \dm}$, we have\footnote{Other positional encoding approaches such as \citep{chen.2021.simple} show similar performances.}
\begin{equation}
  \label{eq:branch.inputs}
  \bH_m^{\rm P}=\bH_m^{(l)}+\hat{\bi}^{\rm L}_m; \quad \bH_m^{\rm S}=\bH_m^{(l)}+\hat{\bi}^{\rm S}_m,
\end{equation}
which are passed into the attention sublayers \eqref{eq:path.attn.weight}--\eqref{eq:path.attn} for message passing.\footnote{We omit the layer indicator $\cdot^{(l)}$ if possible for simplicity.}
The branch outputs $\hat{\bH}^{\rm P}$ and $\hat{\bH}^{\rm S}$ are aggregated by a gating layer that generates the layer output $\hat{\bH}^{(l)}$:
\begin{equation}
  \begin{aligned}
    \hat{\bH}_m^{(l)} = &\g(\hat{\bH}_m^{\rm P}, \hat{\bH}_m^{\rm S})\odot\hat{\bH}_m^{\rm P} + \\
    & \quad (\bm{1} - \g(\hat{\bH}_m^{\rm P}, \hat{\bH}_m^{\rm S}))\odot\hat{\bH}_m^{\rm S}.
  \end{aligned}
\end{equation}
The input of the first layer is the summation of the node embedding and global positional embedding $\bH_m^{(1)} = \e_m + \hat{\bi}^{\rm G}_m$, while the last output $\hat{\bH}^{(N)}$ is fed into a classification layer to predict node labels, assuming the model has $N$ layers in total.

\subsection{Training and Inference}
\label{subsec:training.and.inference}

We use binary cross-entropy as our training objective.
Suppose the predicted \emph{logit} is $\hat{y}$, then the loss at the level of a DOM tree is calculated as
\begin{equation*}
  \ell = -\sum_{m=1}^{|\V|} y_{m}\log \sigma(\hat{y}_{m}) + (1-y_{m})\log \sigma(1-\hat{y}_{m}).
\end{equation*}
During inference, we use $0.5$ as a hard classification threshold for the predicted probability $\sigma(\hat{y})$.

\section{Evaluation}
\label{sec:experiment}

In this section, we first describe a new dataset of interests and their associated webpages, specifically created to benchmark methods for the PT extraction problem. We then evaluate \ours on the same, pitting it against a range of applicable baselines. Finally, we look at the effectiveness of various model components via ablation studies.


\begin{table*}[t!]\small
  \centering
  \begin{threeparttable}
    \begin{tabular}{c|c|c|c|c|c|c|c}
      \toprule
      \multicolumn{2}{c|}{Models} & \dataset-$1$ & \dataset-$2$ & \dataset-$3$ & \dataset-$4$ & \dataset-$5$ & $\bar{\text{F}}_1$ ( precision / recall ) \\
      \midrule
      \multirow{2}{*}{\shortstack{Heuristic\\Methods}}
      & Similarity & 40.12 & 39.14 & 35.84 & 36.55 & 33.80 & 37.09 ( 28.52 / 52.44 ) \\
      & Rules & 56.53 & 62.44 & 56.90 & 59.68 & 58.28 & 58.77 ( 44.20 / \textbf{88.02} ) \\
      \midrule
      \multirow{6}{*}{\shortstack{Supervised\\Methods}}
      & MLP & 66.65 & 66.28 & 66.31 & 74.71 & 61.90 & 67.17 ( 72.11 / 63.38 ) \\
      & BERT-FT & 72.50 & 71.63 & 73.03 & 77.87 & 65.69 & 72.14 ( 68.32 / 76.65 ) \\
      & Graphormer & 71.09 & 81.76 & 75.73 & 66.81 & 69.67 & 73.01 ( 76.61 / 70.89 ) \\
      & GAT & 71.31 & 85.45 & 74.83 & 78.40 & 67.84 & 75.57 ( 77.07 / 74.28 ) \\
      & GCN & 76.13 & 84.07 & \textbf{79.16} & 81.50 & 71.92 & 78.56 ( \textbf{84.44} / 73.57 ) \\
      \cmidrule{2-8}
      & \ours & \textbf{79.65} & \textbf{88.26} & 78.99 & \textbf{82.40} & \textbf{75.35} & \textbf{80.93} ( 84.06 / 77.81 ) \\
      \bottomrule
    \end{tabular}
  \end{threeparttable}
  \caption{
    Test \fone scores on each dataset \dataset-$n$ and the macro-averaged results (in \%).
  }
  \label{tb:main.results}
\end{table*}

\subsection{Experiments}
\label{subsec:experiment.setup}

\paragraph{Dataset}

We constructed a dataset containing \num{95} shopping interests and queried Google for hub pages using automatically selected prompts such as ``[hiking] equipment list'', where ``hiking'' is the SI.
For each SI, we downloaded the top \num{100} returned pages and labeled them with PT nodes using a semi-automatic process. First, we applied simple heuristic rules to create noisy PT labels, based on structure and tag matching. Thereafter, for each SI, we presented roughly \num{5} webpages having a noisy label to a human annotator to further refine the labels. Even so, the dataset is not entirely noise-free given the subjective nature of the labeling process, with many ambiguous cases, such as deciding whether a software such as ``VSCode'' makes a valid product type.
The pages without any positive human label were discarded. This process ultimately resulted in a collection of \num{453} HTML webpages having \num{94167} nodes, among which \num{12548} nodes are positive.
Further details are described in \cref{appsec:dataset.construction.details}.

\paragraph{Setup}

We focus on a zero-shot setup \wrt SIs since our goal is to evaluate various methods on SIs not seen during training. Therefore, we split the collection of webpages \textit{stratified by their associated SIs} (recall that a webpage is assumed to be associated with only one SI) into training ($75 \%$), validation ($10\%$) and test partitions ($15\%$), ensuring that no SI is shared across partitions. As our dataset is small, we randomly split the collection \num{5} times and generated \num{5} distinct datasets, each with the three partitions. This approach is aimed to mitigate the impact of random factors while measuring real model performance. We identify the datasets as \dataset-$n$, where $n\in 1:5$ is the split index.


\paragraph{Baselines}
We consider the following simple to complex methods.
1) \textbf{Heuristic rules} are heuristic functions we manually designed to locate PT nodes from the DOM trees, which were also used to generate the initial, noisy node labels.
2) \textbf{Text similarity} decides whether a node is positive based on the cosine similarity between text and SI embeddings.
3) \textbf{Fine-tuned BERT} (BERT-FT) fine-tunes a BERT-base model to independently classify each tree node based on its text.
4) \textbf{Multilayer perceptron} (MLP) also classifies each node independently, but with fixed BERT text embeddings followed by a set of FFN layers.
5) Graph neural networks (GNNs) propagate node semantics throughout the graph by aggregating neighboring node embeddings.
GNN family has many variances, and we focus on \textbf{GCN} and \textbf{GAT}.
6) \textbf{Graphormer} \citep{Ying.2021.Graphormer}, designed for molecular graphs, adds special encodings and attention masks to the Transformer model.
Please see \cref{appsubsec:imp.details.baselines} for implementation details.

\paragraph{Metrics}

We evaluate each model with the \fone scores corresponding to each split \dataset-$i$ and the macro-averaged \fone score $\bar{\text{F}}_1 = \frac{1}{5}\sum_{n=1}^5 {\text{F}_1}_n$ with the corresponding macro precision and recall.

All trainable methods are equipped with early stopping techniques based on validation \fone scores.
To further reduce the influence of random factors without increasing training pressure, we store \num{5} snapshots of the models that perform the best on the validation dataset during training.
During the test, we predict \num{5} sets of labels from the model snapshots and use the majority-voted labels as the final model predictions.
It can be regarded as a simplified model ensemble method often used to improve model robustness \citep{Dong.2020.ensemble}.

\begin{figure}[tbp]
  \centering{
    \subfloat[\fone \vs DOM tree depth]{
      \label{subfig:dom.f1}
      \includegraphics[height=1.13in]{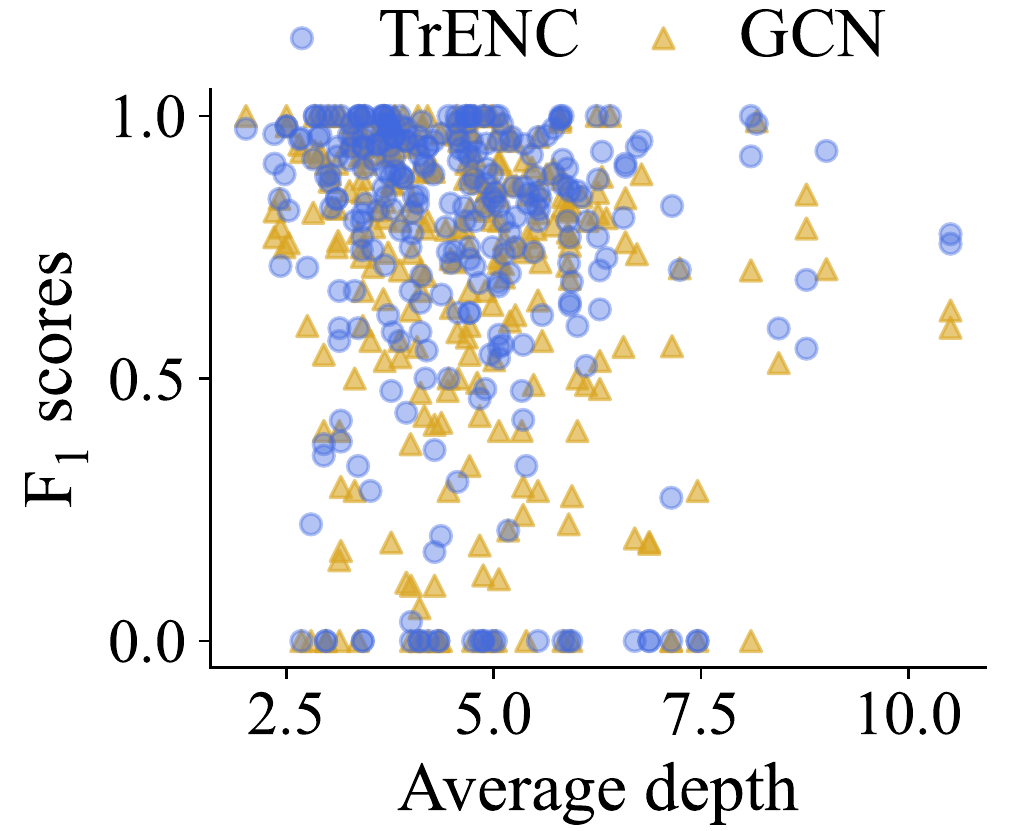}
    }
    \subfloat[Average \fone \vs depth level]{
      \label{subfig:quantile.f1}
      \includegraphics[height=1.13in]{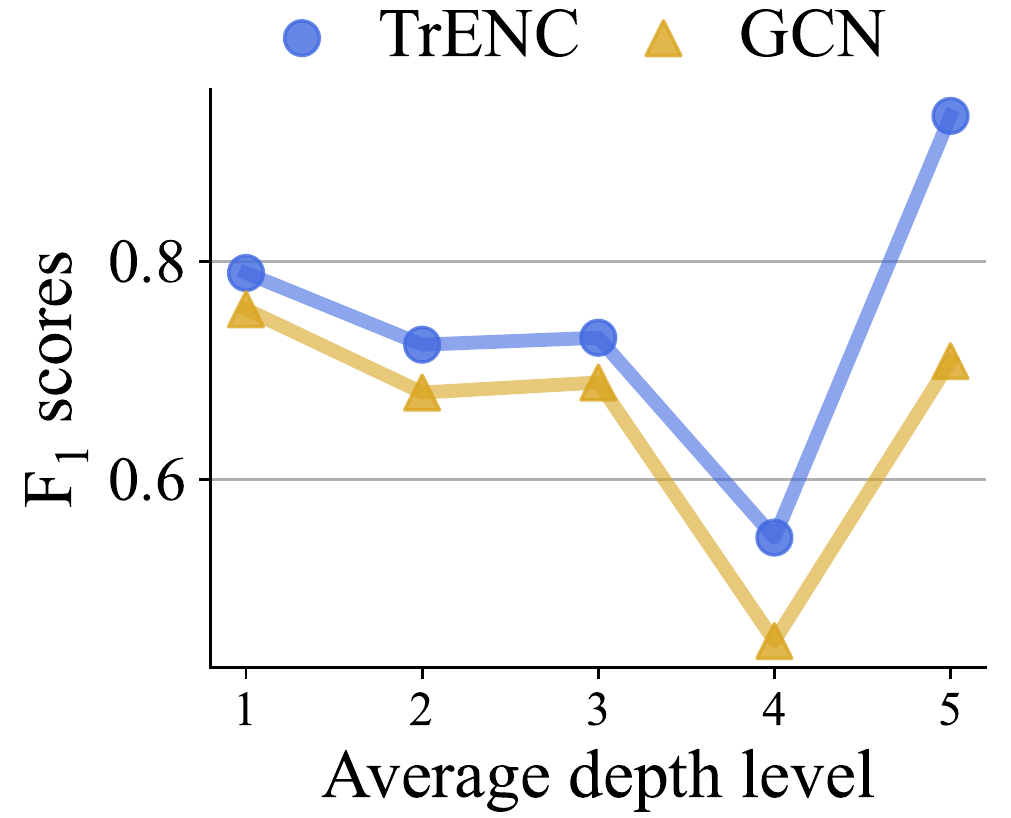}
    }
  }
  \caption{
    Test \fone scores against the DOM tree depths.
  }
  \label{fig:depth.performance}
\end{figure}

\subsection{Main Results}
\label{subsec:results}

Table~\ref{tb:main.results} shows the results of our comparative evaluation.
As seen, \ours outperforms all methods, exceeding the strongest baseline, GCN, by a margin of $2.37$ absolute \fone on average.
Considering the small size of our datasets, it is not surprising that the test \fone scores have relatively large variation across different data splits, as the correlation of data distributions of the training and test sets is susceptible to random factors.
Nonetheless, \ours achieves the best performance on $4$ out of $5$ splits as well as exceeds by a good margin on average, which strengthens the confidence of evaluation.
Surprisingly, Graphormer underperforms GNN models and barely outperforms BERT-FT, a model that treats nodes independently without considering the tree structures.
It indicates that models designed for other graphs such as molecular graphs are not directly applicable to our case.
Instead of helping, the features Graphormer emphasizes \emph{prevent} the model from learning a reasonable representation of the DOM tree.
Table~\ref{tb:main.results} also shows that the cosine similarity between SI and PT embeddings does not present a good performance.
This is not unexpected as SI and PTs are not usually semantically similar, making it a sub-optimal way to directly compare their embeddings.

We also compare \ours with GCN at varying levels of DOM tree complexity.
Figure~\ref{subfig:dom.f1} shows tree-level \fone scores of each DOM tree against its depth, which is the average depth of its nodes $\frac{1}{|\V|} \sum_{m=1}^{|\V|} i_{m}^{\rm L}$ and roughly echos the tree complexity.
Figure~\ref{subfig:quantile.f1} divides the depth equally into \num{5} levels and presents the average \fone for each level.
As seen, \ours has better overall performance than GCN at all depths.
In addition, the gap between \ours and GCN increases when the tree is deeper, which indicates that \ours can better encode complex trees due to the global message-passing ability of the self-attention mechanism.

\begin{table}[t!]\small
  \centering
  \begin{threeparttable}
    \begin{tabular}{c|c|c|c}
      \toprule
      \multicolumn{2}{c|}{Models} & Average \fone & Gap \\
      \midrule
      \multicolumn{2}{c|}{\ours} & 80.93 & - \\
      \midrule
      \multirow{3}{*}{\shortstack{Input\\features}}
      & w/o SI & 79.83 & 1.10 \dn  \\
      & w/o tag & 78.82 & 2.11 \dn \\
      & w/o text & 57.98 & 22.95 \dn \\
      \midrule
      \multirow{5}{*}{\shortstack{Model\\components}}
      & w/o gating & 80.39 & 0.54 \dn \\
      & Transformer & 78.44 & 2.49 \dn \\
      & w/o pos emb & 79.60 & 1.33 \dn \\
      & w/o pth attn & 78.98 & 1.95 \dn \\
      & w/o sbl attn & 78.04 & 2.89 \dn \\
      \midrule
      \multirow{3}{*}{\shortstack{Sequence\\encoding}}
      & BERT-large & 79.36 & 1.57 \dn \\
      & RoBERTa & 78.54 & 2.39 \dn \\
      & Sentence-BERT & 74.73 & 6.20 \dn \\
      \bottomrule
    \end{tabular}
  \end{threeparttable}
  \caption{
    Ablation study \fone scores (in \%).
    ``Pth'' is short for path; ``sbl'' is short for sibling; and ``pos emb'' represents positional embeddings.
  }
  \label{tb:ablation}
\end{table}

\subsection{Ablation Studies}
\label{subsec:ablation}

We ablate input features and model components from \ours to understand their effectiveness.
Table~\ref{tb:ablation} shows the ablation results.

\paragraph{Input Features}

As seen, although removing any input feature (\cref{subsec:input.features}) impairs the model performance, text sequence is the most critical feature for \ours.
We further notice that without text sequence, \ours performs quite close to the heuristic rules that utilize very limited lexical features (Table~\ref{tb:main.results}).
This may indicate that \ours exhausts the structural information available in a DOM tree.

Although not as significant as text sequences, incorporating SIs and tags does enhance the model performance.
Injecting SIs turns the model's attention to their correlation with PTs.
But such improvement is limited as the correlation is not strong, as discussed in \cref{subsec:results}.

\paragraph{Model Components}

We investigate the functionalities of model components by removing them separately.
The Transformer model discards edges $\E$ and treats the tree as a linearized sequence of nodes arranged by their global positional indices $i^{\rm G}$.
Although it learns certain structural dependencies, as indicated by its advance in comparison with MLP (Table~\ref{tb:main.results}), missing explicit edge knowledge still affects the model's judgment.

The results also show that path attention, sibling attention and positional encodings all contribute to better tree encoding.
The row ``w/o pos enc'' removes the level and sibling encodings $\bi^{\rm L}, \bi^{\rm S}$ but keeps the global encoding $\bi^{\rm G}$.
Without $\bi^{\rm L}$ and $\bi^{\rm S}$, the model cannot properly identify the hierarchy and sibling order between nodes and therefore performs worse.
Compared to path attention, sibling attention demonstrates a higher importance in context understanding, even though removing path attention means a node no longer has access to any other nodes from the tree.

\paragraph{Sequence Encoding}
In our implementation, we use the uncased BERT-base model with $d^{\rm BERT}=768$ as our encoders for sequence embeddings $\e$ and concept embeddings $\bc$.\footnote{\href{https://huggingface.co/bert-base-uncased}{https://huggingface.co/bert-base-uncased}}
The embeddings are fixed during the training process.

We also test other pre-trained language models, including BERT-large, RoBERTa \citep{Liu.2019.RoBERTa} and Sentence-BERT \citep{Reimers.2019.Sentence.BERT}, which is designed for comparing the sequence similarities and claims better sentence embedding performance than BERT.
However, Table~\ref{tb:ablation} shows that none outperforms BERT-base \citep{Devlin.2019.BERT}.
The reason might be the incompatibility of their training corpus and objective to our task.
The results indicate that choosing an encoding model is vital to have good performance.

\begin{table}[t!]\small
  \centering
  \begin{threeparttable}
    \begin{tabular}{c|c|c}
      \toprule
      & SI & Node text sequence \\
      \midrule
      \multirow{3}{*}{\shortstack{FP}}
      & at-home-spa & Esthetics or Skin Care \\
      & hiking & Merrell Overlook Tall 2 WP Boot \\
      & running & Credit card \\
      \midrule
      \multirow{2}{*}{\shortstack{FN}}
      & fishing & Rods for River Fishing \\
      & canoeing & Water bottle - 1 litre is good \\
      \bottomrule
    \end{tabular}
  \end{threeparttable}
  \caption{
    Examples of common mistakes made by \ours.
    FP/FN indicates false positives/negatives.
  }
  \label{tb:mistake.study}
\end{table}

\subsection{Case Studies on Classification Mistakes}
\label{subsec:classification.mistakes}
Table~\ref{tb:mistake.study} shows a few false positive (FP) and false negative (FN) examples to illustrate certain text sequence patterns where \ours fails.
As seen from FP cases, \ours either struggles to determine whether it is a broad PT category ($1^{\rm st}$ row), has challenges discerning a PT from a specific product ($2^{\rm nd}$ row), or makes mistakes when unavoidable non-purchasable items are mentioned on the page along with other valid PTs ($3^{\rm rd}$ row).
From the FN cases, we conjecture that long descriptions may overwhelm the textual semantics and deviate its embedding, thereby preventing \ours predict correctly ($4^{\rm th}$ \& $5^{\rm th}$ rows).
The reason might be that \ours have a stronger dependency on the node semantics than the structure, which is also indicated by the ablation results, and properly balancing the conditional terms may mitigate this issue.

\section{Conclusion}
\label{sec:conclusion}

In this paper, we consider a new problem of extracting product types from the Web pages that are relevant to broad shopping interests such as camping.
We model the problem as a node classification task and propose \ours, a Transformer encoder-based model that leverages unique characteristics of DOM trees to perform product type extraction.
In addition to the node-level signals including HTML tags, text sequences and shopping interest semantics, \ours design path and sibling attention mechanisms based on DOM tree's ancestor-descendant and sibling relations.
Together with the tree-based positional embeddings, the structural attention mechanisms promote the tree architecture understanding and make the classification more effective.
Zero-shot experiments on a new dataset \ours containing \num{95} shopping interests and \num{453} pages show that \ours outperforms the baseline graph encoding models.
This work pushes the frountier of researches of a more organized and intuitive result recommendation for middle-funnel customers.


\section*{Limitations}

Apart from the issues mentioned in \cref{subsec:classification.mistakes}, another limitation of \ours is that it does not integrate any pre-training process such as BERT, which is effective in increasing the language understanding ability and adopted by previous works focusing on token-level classification tasks \citep{Wang.2022.WebFormer, Deng.2022.DOM-LM}.
Two factors lead to this decision.
First, we use DOM nodes instead of tokens as the classification object and focus on relations between nodes rather than tokens.
As the node text sequence is a composition of an arbitrary number of tokens, adopting the conventional masked language modeling (MLM) training objective \citep{Devlin.2019.BERT} seems impractical since there is \emph{no direct mapping} from an embedding vector, one-hot encoded or not, to a sentence.
The second reason is simply that we do not possess the corpus or computation resources for model pre-training.
In fact, we expect a properly designed pre-training scheme to bring better node semantics representation and SI-PT relation modeling.
It is an interesting topic and deserves further study.

\section*{Acknowledgments}

This work was supported in part by Amazon.com Services LLC, NSF IIS-2008334, IIS-2106961, and CAREER IIS-2144338.

We would like to thank Xian Li, Binxuan Huang, Chenwei Zhang, Yan Liang, and Jingbo Shang for their insightful advice on this work.

\bibliographystyle{acl_natbib}
\bibliography{references}

\clearpage

\appendix

\section{Dataset Details}
\label{appsec:dataset.construction.details}

\subsection{Dataset Construction}

We build \dataset to realize a quantitative analysis of different PT extraction methods.
\dataset is a collection of hub pages relevant to a set of pre-defined SIs.
Its construction process mainly consists of \num{5} steps: 1) defining SIs; 2) crawling hub pages; 3) processing HTML documents; 4) labeling documents; and 5) splitting data points.

\paragraph{Defining Shopping Interests}
As the first step, we establish a set of SIs through brainstorming.
Particularly, we focus on popular activities, sports, hobbies and special events.
Please check Table~\ref{apptb:sis} and \ref{apptb:sis.ct} for a complete list of SIs.

\paragraph{Crawling Hub Pages}
The hub pages are the webpages, each providing PTs related to a specific SI.
Due to the variety of SIs, it is infeasible to focus on one or several websites for hub page collection.
For example, a website specializing in sports will not provide information on ``sewing'' with a high chance and vice versa.
In addition, gathering information from different websites may eliminate the bias probably existing in one website, according to the law of large numbers.

Considering this situation, we take advantage of Google Search with a simple query selector to locate the hub pages.
Each SI $C$ is combined with suffices ``equipment list'', ``supply list'', ``tool list'' and ``checklist'' before being fed into the search engine for querying.
The system selects the combination with the largest number of results, whose top-\num{100} query results are saved for later usage.
We keep only the HTML pages and discard other documents such as PDFs or CSVs, so the actual number of saved documents may vary.

\begin{figure}[!t]
    \centerline{\includegraphics[width = 0.48\textwidth]{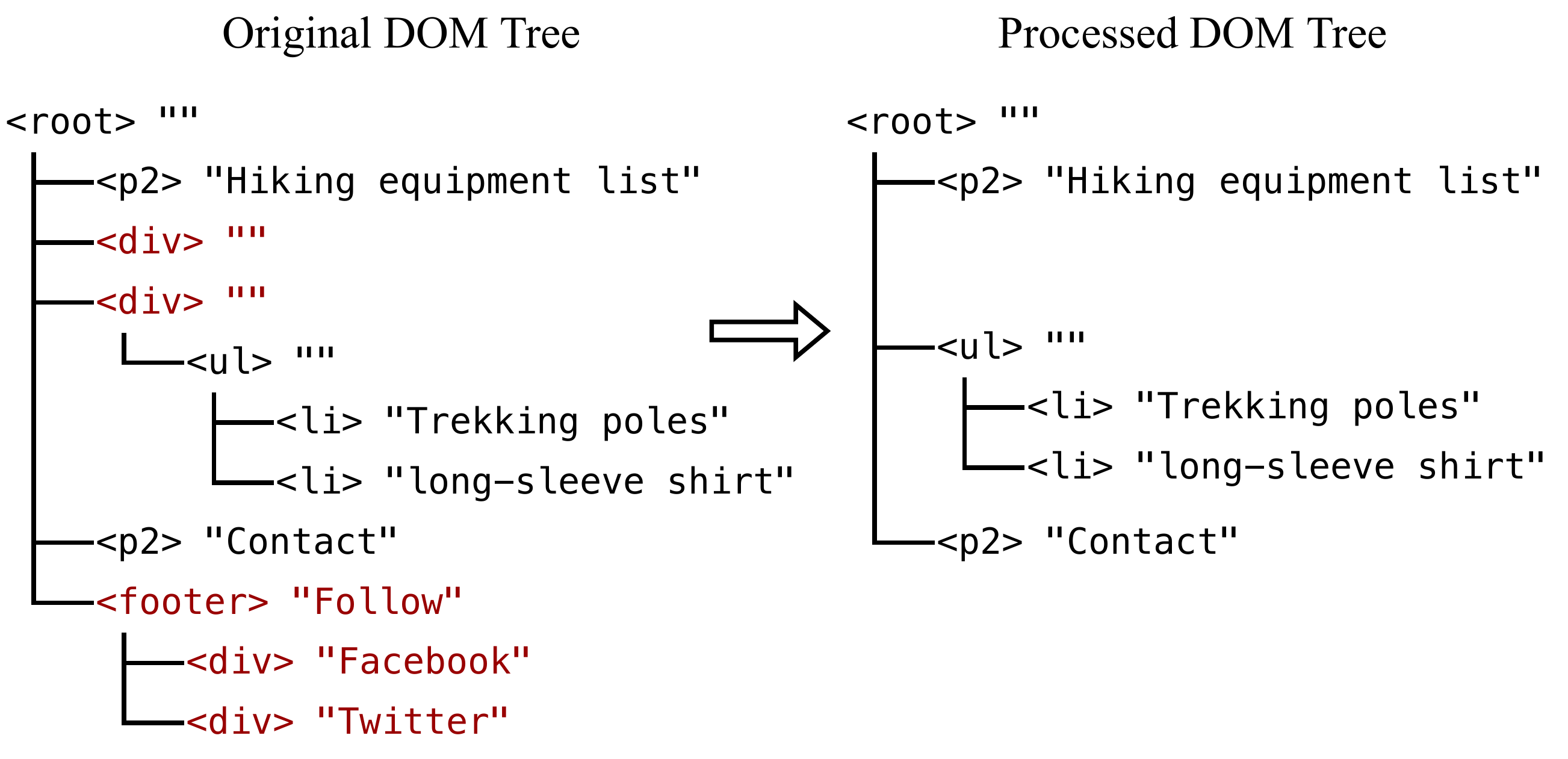}}
    \caption{
        An example of HTML processing.
    }
    \label{appfig:a1-dom-trim}
\end{figure}

\paragraph{Processing HTML Documents}
This step aims to simplify the DOM tree structure to facilitate PT extraction.
The RAW DOM tree is complicated with decorative and supporting scripts irrelevant to the content, which easily submerges the useful information we want to extract and decreases the false positive rate.
We prune the trees by removing all headers, footers, and leaf nodes with empty text sequences.
Then, we replace the nodes with only one child by their children to decrease the tree depth.
To reduce the tree depth, we delete the nodes with only one child and then connect their children with subsequent subtrees directly with their parents.
The process is illustrated in Figure~\ref{appfig:a1-dom-trim}.
Experiments show that this HTML processing strategy successively simplifies the DOM structure without sacrificing any targeted content.

\paragraph{Labeling Documents and Splitting Data Points}

These two steps are sufficiently discussed in \cref{subsec:experiment.setup} as will not be repeated.
The only supplement is that the heuristic method used for initializing the noisy labels and compared in Table~\ref{tb:main.results} is empirically developed.
We omit its discussion since it is complex and not the focus of this paper.
The detailed dataset splits are presented in Table~\ref{apptb:sis} and \ref{apptb:sis.ct}.

\begin{figure}[!t]
    \centerline{\includegraphics[width = 0.48\textwidth]{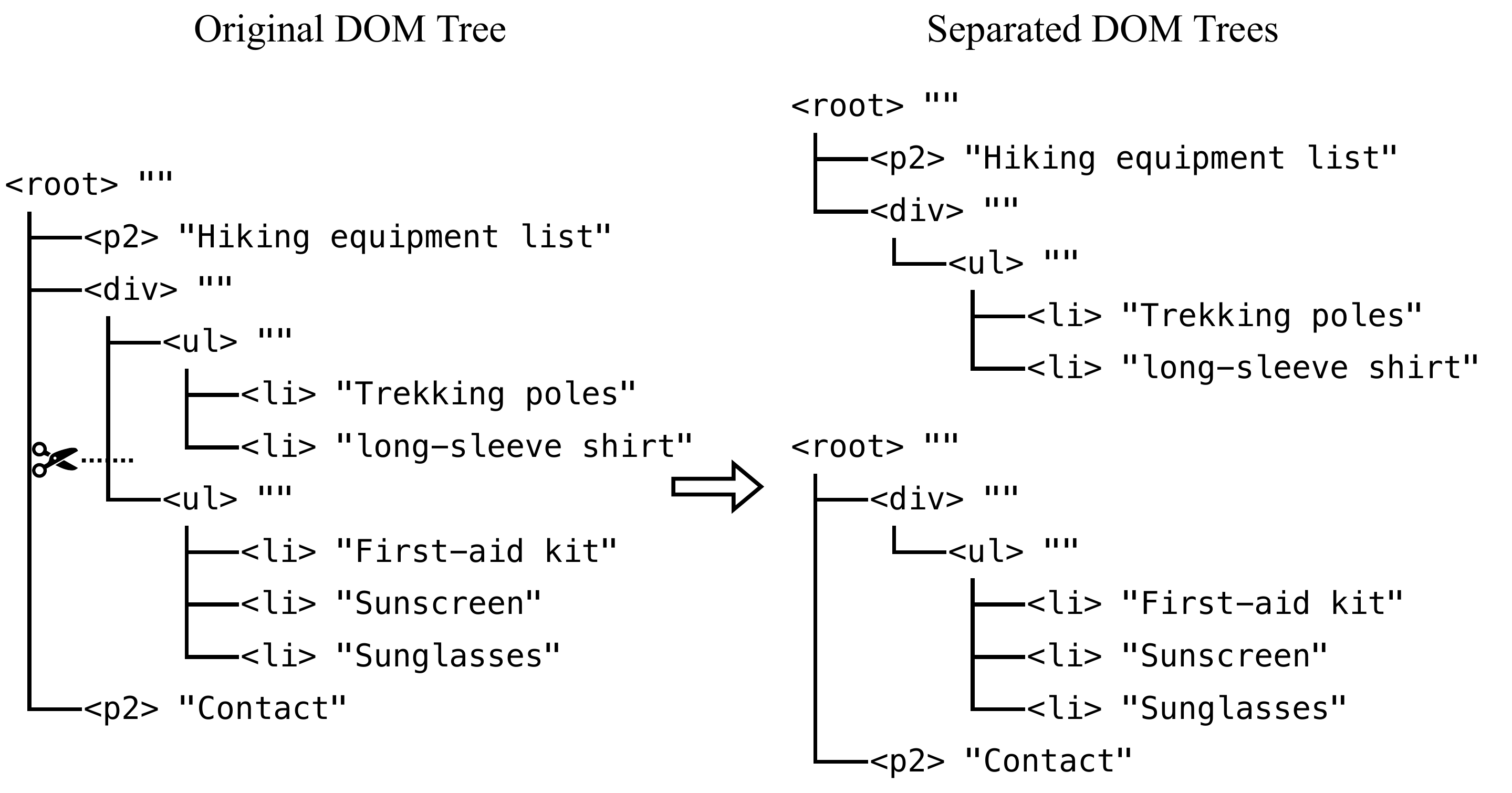}}
    \caption{
        An example of separating a DOM tree.
    }
    \label{appfig:a1-dom-cut}
\end{figure}

\paragraph{Data Processing for Transformers}
One limitation of the Transformer models such as BERT and \ours is that they need to set a constraint to the length of the input sequence $|\V|$ since the complexity of the self-attention mechanism is $\calO(|\V|^2)$ and easily explodes when $|\V|$ is too large.
Considering this drawback, for the node Transformers including Graphormer and \ours, we set \num{512} as the maximum size of a DOM tree and split those that exceed this size.
In addition, we guarantee that each split tree has \num{64} nodes at minimum.
Figure~\ref{appfig:a1-dom-cut} shows an example of the separation process.

\subsection{Dataset Statistics}

We present the dataset statistics in Table~\ref{apptb:dataset.statistics}.
DOM trees are larger than molecular graphs but significantly smaller than knowledge graphs.

\begin{table}[t!]\small
  \centering
  \begin{threeparttable}
    \begin{tabular}{c|c}
      \toprule
      Attribute & Value \\
      \midrule
      \# Shopping Interests & \num{95} \\
      \# DOM Trees & \num{453} \\
      \# Total Nodes & \num{94167} \\
      \# Leaf Nodes & \num{70161} \\
      \# Positive PT Nodes & \num{12548} \\
      \midrule
      Average \# Nodes per Tree & \num{207.87} \\
      Maximum \# Nodes in a Tree & \num{2748} \\
      Minimum \# Nodes in a Tree & \num{19} \\
      Median \# Nodes in a Tree & \num{156} \\
      Average Tree Depth & \num{7.06} \\
      Maximum Tree Depth & \num{18} \\
      Minimum Tree Depth & \num{3} \\
      Median Tree Depth & \num{7} \\
      \midrule
      Average \# Trees per SI & \num{4.77} \\
      Average \# Nodes per SI & \num{991.23} \\
      Maximum \# Nodes for an SI & \num{3050} \\
      Minimum \# Nodes for an SI & \num{363} \\
      Median \# Nodes for an SI & \num{935} \\
      \bottomrule
    \end{tabular}
  \end{threeparttable}
  \caption{
    Dataset statistics.
  }
  \label{apptb:dataset.statistics}
\end{table}

\subsection{Labeling Quality}

\begin{figure}[tbh]
    \centerline{\includegraphics[width = 0.45\textwidth]{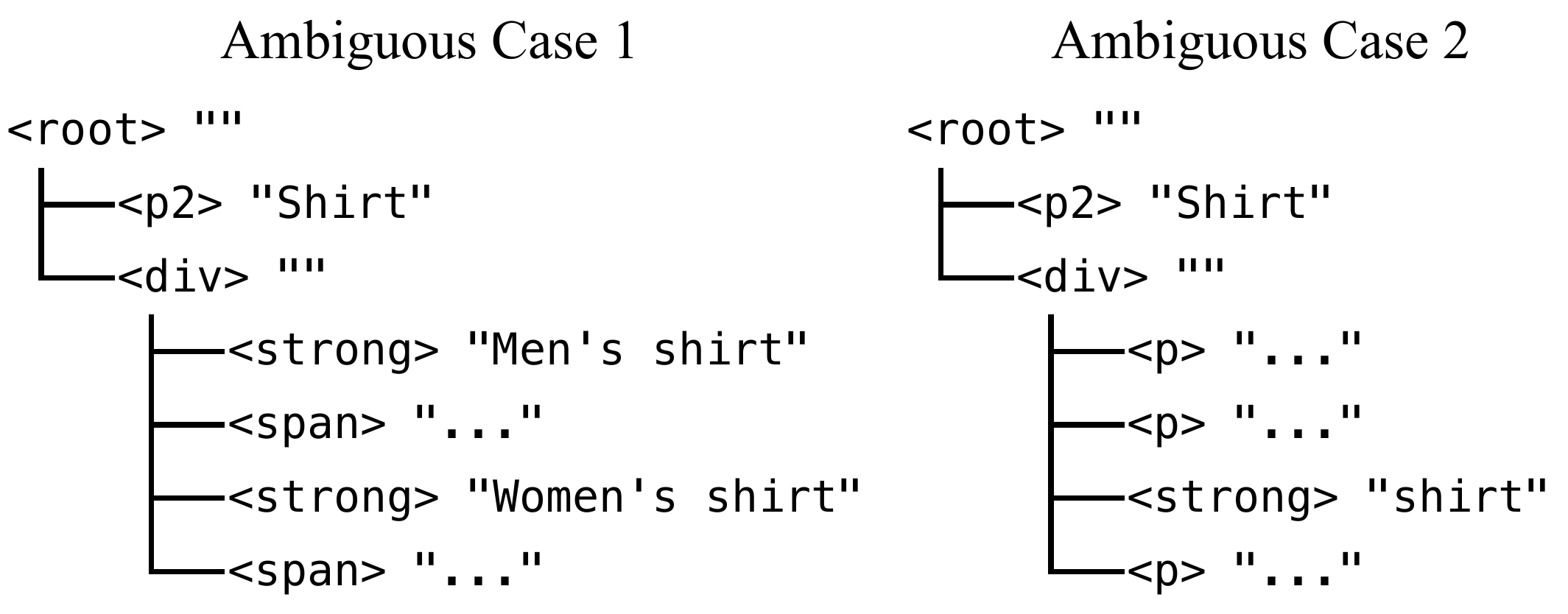}}
    \caption{
        Examples of typical ambiguous annotation cases.
        In case 1, annotators may regard ``men's shirt'' and ``woman's shirt'' as negative as they are subcategories of the PT ``shirt'';
        in case 2, the latter ``shirt'' may be regarded as negative as it is a repetition surrounded by long descriptive sentences.
    }
    \label{appfig:a1-ambiguous}
\end{figure}

The dataset is labeled by one individual as the task is straightforward.
To investigate the labeling quality, we randomly select \num{25} DOM trees, removing their original labels and presenting them to \num{2} individuals for re-labeling.
Table~\ref{apptb:annotation.quality} presents the statistics and results.
It shows that our labeling quality is decent despite some inevitable disagreements on ambiguous cases, as exampled in Figure~\ref{appfig:a1-ambiguous}.

\begin{table}[tbh]\small
  \centering
  \begin{threeparttable}
    \begin{tabular}{c|c}
      \toprule
      Attribute & Value \\
      \midrule
      \# DOM Trees & \num{25} \\
      \# Total Nodes & \num{5938} \\
      \# Positive PT Nodes & \num{683} \\
      \# Disagreement & \num{87} \\
      \midrule
      \# Fleiss' $\kappa$ & \num{98.53} \\
      \bottomrule
    \end{tabular}
  \end{threeparttable}
  \caption{
    Annotation quality investigation.
  }
  \label{apptb:annotation.quality}
\end{table}

\subsection{Data Usage}

All Web pages used by \dataset are included in the Common Crawl repository.\footnote{\href{https://commoncrawl.org/}{https://commoncrawl.org/}}
They are intended to provide information on a topic or interest, so consistent with that idea, we labeled the product types on each page.
The labels do not contain any personally identifiable information.
We are making the annotated dataset available to encourage further research on the product type extraction problem.

The \dataset dataset is licensed under the Creative Commons Attribution 4.0 International License.
To view a copy of this license, visit \href{http://creativecommons.org/licenses/by/4.0/}{http://creativecommons.org/licenses/by/4.0/}.

\section{Implementation Details}
\label{appsec:implementation.details}

\subsection{\ours Hyper-Parameters}
\label{appsubsec:imp.details.treeformer}

We set the model dimensionality $\dm = 128$ and the number of \ours layers $N=12$.
Each attention branch has \num{4} attention heads, and the single-head attention dimensionality $d_k=32$.
The feed-forward layer above the attention layer (Figure~\ref{fig:s4-model-architecture}) first maps the features from $\dm$ to a $512$-dimensional latent space and then maps it back.
The classification layer consists of \num{2} FFN sublayers that first downscale the \ours layer output to \num{16}-dimensional and then to the \num{1}-dimensional output logits $\hat{y}$.
We use the same activation functions and dropout strategy as described in \citep{Vaswani.2017.Transformer}.
Our experiments show that the performance remains similar when we use $6$ or $8$ as the number of heads or use model dimensionality $\dm = 512$.

We train the model using $10^{-4}$ as the peak learning rate of the AdamW optimizer \citep{Loshchilov.2019.AdamW} with linear scheduler with $0.1$ warm-up ratio.
The batch size is \num{8} and the random seed is \num{42}.
We do not take multiple runs for each model on each dataset as our dataset and evaluation strategies (\cref{subsec:experiment.setup}) can minimize the impact of random factors.
Using another random seed (\num{0}) only changes the $\bar{\text{F}}_1$ scores of \ours and GCN by $0.03$ and $0.05$, respectively.
The model is implemented with the ``Transformers'' library \citep{wolf.2020.transformers} in PyTorch \citep{Paszke.2019.Transformers}.
The hyper-parameters not mentioned above keep their default values.

\subsection{Baseline Methods}
\label{appsubsec:imp.details.baselines}

\paragraph{Text Similarity}

We adopt the same approach as described in \cref{subsec:input.features} with the uncased BERT-base model to generate the text sequence embedding $\e_m$ of each node $V_m$ and the concept embeddings $\bc$.
Then, we compute their cosine similarity through
\begin{equation*}
  {\rm sim}_m \in (0,1) = \frac{\e_m^\tr \bc}{\|\e_m\| \|\bc\|}.
\end{equation*}
We decide the classification threshold by exhausting possible values with $0.01$ interval within $(0,1)$ and select the one that gives the largest \fone score.
Notice that this threshold searching method is only applied to the text similarity baseline.
Others take a constant threshold $0.5$, as described in \cref{subsec:training.and.inference}.

\paragraph{BERT-FT}

BERT-FT classifies each node $V_m$ independently by fine-tuning the uncased BERT-base model with the sequence classification task.
The model input is the combination of the sequence $\calS_m$ and the concept $C$, \ie, ``\texttt{[CLS]} $\calS_m$ \texttt{[SEP]} $C$ \texttt{[SEP]}''.
It does not consider the tag $t_m$.
We append a one-layer FFN to the embedding corresponding to the \texttt{[CLS]} token to map it to a $1$-dimensional logit.
The training objective is minimizing the binary cross-entropy.

\paragraph{MLP}

MLP can be considered as a \ours model without \ours layers.
In other words, it directly feeds the node embeddings $\e$ (\cref{subsec:input.features}) into the classification layer (Figure~\ref{fig:s4-model-architecture}) without considering any inter-dependencies between nodes.
We increase its classification layer depth until the validation \fone stops improving for a fair comparison.

\paragraph{GNNs}

Similar to MLP, GNN models substitute the \ours layers in the \ours model with the GCN and GAT layers, respectively.
The GNN layers are implemented with the ``PyTorch Geometric'' library \citep{Fey.2019.PyG}.
The number of GNN layers is fine-tuned according to the validation performance.

\paragraph{Graphormer}

We take the original implementation of \citet{Ying.2021.Graphormer} and keep all model components.\footnote{\href{https://github.com/microsoft/Graphormer}{https://github.com/microsoft/Graphormer}.}
The differences are that we initialize the node features with node embeddings $\e$ instead of atom categories, and we train the model with node classification instead of graph classification.
We keep its scheme for encoding edges but introduce only one edge category representing the ancestor-descendant relationship.

\begin{table}[tbh]\small
  \centering
  \begin{threeparttable}
    \begin{tabular}{c|c|c|c|c|c}
      \toprule
      \multirow{2}{*}{SI}
      & \multicolumn{5}{c}{\dataset-$n$} \\
      & \num{1} & \num{2} & \num{3} & \num{4} & \num{5} \\
      \midrule
      3d-printing & tr & tr & tr & tr & tr \\
      airsoft-paintball & tr & tt & tr & tt & tr \\
      archery & tr & tr & tt & vl & tr \\
      astronomy & tr & tr & tr & tr & tr \\
      at-home-fitness & tr & vl & tr & tt & tt \\
      at-home-spa & tt & tr & tr & vl & tr \\
      badminton & tr & vl & tr & tr & vl \\
      baking & tr & tr & vl & tr & tr \\
      bartending & vl & tr & tr & tr & tr \\
      baseball & tr & tr & tr & tr & tr \\
      basketball & tr & tr & vl & tr & tt \\
      billiards-pool & tt & vl & vl & tt & tr \\
      bird-watching & tr & tt & tr & tt & tr \\
      boating & tr & tr & tt & tr & tr \\
      bowling & tr & tt & tt & tr & vl \\
      boxing & tr & tr & tr & tr & tr \\
      calligraphy & tr & tr & vl & tr & vl \\
      camping & tr & tr & tr & tr & tr \\
      candle-making & tr & tr & tr & tt & vl \\
      canoeing & tr & tt & tr & tr & tr \\
      cheerleading & tr & tr & tr & tr & tr \\
      cleaning & tr & tr & tr & tr & tr \\
      climbing & tt & tr & tr & tr & vl \\
      coffee & tr & tr & tr & tr & tr \\
      comics-manga & tr & tr & tr & tr & tr \\
      content-creation & tt & vl & tt & tr & vl \\
      cricket & tr & tr & vl & tt & tr \\
      crossfit & vl & tr & tr & tr & tr \\
      cycling & tr & tr & tt & tr & tr \\
      digital-art & tr & tr & tr & vl & tr \\
      diy-home-improvement & tr & tr & tr & tr & tr \\
      dj & tr & tr & tr & tr & tr \\
      drag-queen & tr & tr & tr & tr & tt \\
      drawing-and-sketching & tr & tr & tt & tr & tr \\
      fencing & tr & tr & tt & tr & tr \\
      field-hockey & tr & vl & vl & tr & tr \\
      fishing & tt & vl & tr & tr & tr \\
      floral-arranging & tr & tr & tr & tr & tr \\
      football & vl & tr & tr & tr & tr \\
      gaming & tt & tr & tr & tr & tr \\
      gardening & tr & tt & vl & vl & vl \\
      golfing & tr & tr & tr & tr & tr \\
      gymnastics & vl & tr & tr & tr & tr \\
      hair-care & tr & tr & vl & vl & tt \\
      hiking & tt & tr & tr & tr & tr \\
      hockey & tr & vl & tr & tr & tr \\
      home-entertainment & tt & vl & tr & tr & tt \\
      home-schooling & tr & tr & tr & tr & tr \\
      horse-riding & tr & tr & tr & tr & tt \\
      \bottomrule
    \end{tabular}
  \end{threeparttable}
  \caption{
    Shopping interests and their splits in each dataset.
    ``Tr'', ``vl'' and ``tt'' represent ``training'', ``validation'' and ``test'' respectively.
  }
  \label{apptb:sis}
\end{table}

\begin{table}[t!]\small
  \centering
  \begin{threeparttable}
    \begin{tabular}{c|c|c|c|c|c}
      \toprule
      indoor-plants & tr & tr & tt & vl & tr \\
      interior-design & tr & tr & tr & vl & tr \\
      kayaking & tr & tr & tr & tr & vl \\
      knitting & tr & tt & tr & tr & tr \\
      lacrosse & tr & tt & tt & tr & tr \\
      leathercraft & vl & tt & tr & tr & tr \\
      makeup & tr & tr & tr & tr & tt \\
      model-trains & tr & tt & tr & tr & tr \\
      music-production & vl & tr & tr & tr & tr \\
      nails & tr & tr & tt & tr & tr \\
      painting & tr & tr & tr & tr & tt \\
      paper-crafting & tr & tr & tr & tr & tr \\
      parenting & tr & tr & tr & tr & tr \\
      party-planning & tr & tr & tr & tr & tt \\
      pet & tt & tr & tr & tt & tr \\
      pilates & tr & tr & tr & tr & tr \\
      pottery & tt & tr & tr & tr & tr \\
      rugby & tr & tr & tr & tt & tr \\
      running & tr & tt & tr & tr & tr \\
      sailing & tr & tr & tr & tr & tr \\
      scrapbooking & tr & tr & tr & tr & tr \\
      scuba-diving & tr & tr & tt & tr & tt \\
      sewing & tr & tr & tr & tr & tr \\
      skating & tr & tr & tr & tr & tr \\
      skiiing & tt & tr & tr & tr & tr \\
      skin-care & vl & tr & vl & tr & tr \\
      smart-home & tr & tr & tr & tr & tr \\
      snowboarding & vl & tr & tr & tr & tr \\
      soap-making & vl & tr & tr & tr & tt \\
      soccer & tr & tr & tr & tt & vl \\
      softball & tr & tr & tr & tr & tr \\
      storage-and-organization & tr & tt & tr & tr & tt \\
      student-dorm & tr & tt & tr & tr & tr \\
      surfing & tr & tr & tr & vl & tr \\
      swimming & tr & tr & tr & tt & tr \\
      table-tennis & tr & tr & tt & tr & tt \\
      teaching & tt & vl & tr & tt & tr \\
      tennis & tt & tr & tr & tr & tr \\
      travel & tr & tr & tt & tr & tr \\
      volleyball & tr & tr & tr & tr & tr \\
      weaving-and-spinning & tt & vl & vl & vl & tr \\
      wedding & tr & tt & tr & tt & tr \\
      wine & tr & tr & tr & vl & vl \\
      work-from-home & vl & tt & tr & tt & tt \\
      wrestling & tr & tr & tr & tr & tr \\
      yoga & tr & tr & tt & tt & tr \\
      \bottomrule
    \end{tabular}
  \end{threeparttable}
  \caption{
    SIs and splits (cont.).
  }
  \label{apptb:sis.ct}
\end{table}

\end{document}